\documentclass[9pt,conference]{IEEEtran}
\usepackage{dcase2026}

\usepackage{bm} 

\usepackage{dcase2026,amsmath,graphicx,url,times,booktabs, tabularx}

\title{Confidence-Aware Aggregation for Multimodal Audio Reasoning}

\name{Michel Olvera,
      Paraskevas Stamatiadis,
      Gaël Richard}
\address{LTCI, Télécom Paris, Institut Polytechnique de Paris, France \;
}

\begin{document}

\maketitle

\begin{abstract}
Large Audio Language Models (LALMs) enable generative reasoning over complex acoustic scenes but often exhibit prediction instability. We show that this variability can serve as a reliability metric via inference-time self-consistency and confidence-aware aggregation. Using a log-probability–based scoring framework, candidate answers are weighted by model likelihoods, leveraging reasoning diversity and self-certainty without additional training. Experiments on four LALMs, including Qwen3-Omni and Audio Flamingo 3, demonstrate that confidence-aware aggregation improves accuracy and calibration, with additive methods like Borda voting outperforming classical Majority Vote, especially under high disagreement. This approach transforms LALM variability into a source of reliability for scalable and robust multimodal audio understanding. Project code is available at \url{https://github.com/michelolzam/lalm-self-consistency}.
\end{abstract}


\section{Introduction}
\label{sec:intro}

Large Audio Language Models (LALMs) couple audio encoders with large language model backbones to enable generative reasoning over complex acoustic scenes. Despite substantial advances in multimodal representation learning, model scaling, and post-training strategies, reasoning-centric evaluations of state-of-the-art LALMs reveal a persistent mismatch between single-shot predictions and reliable decision making, with high average accuracy often concealing prediction instability.

\begin{figure}[t]
  \centering
  \includegraphics[width=\linewidth]{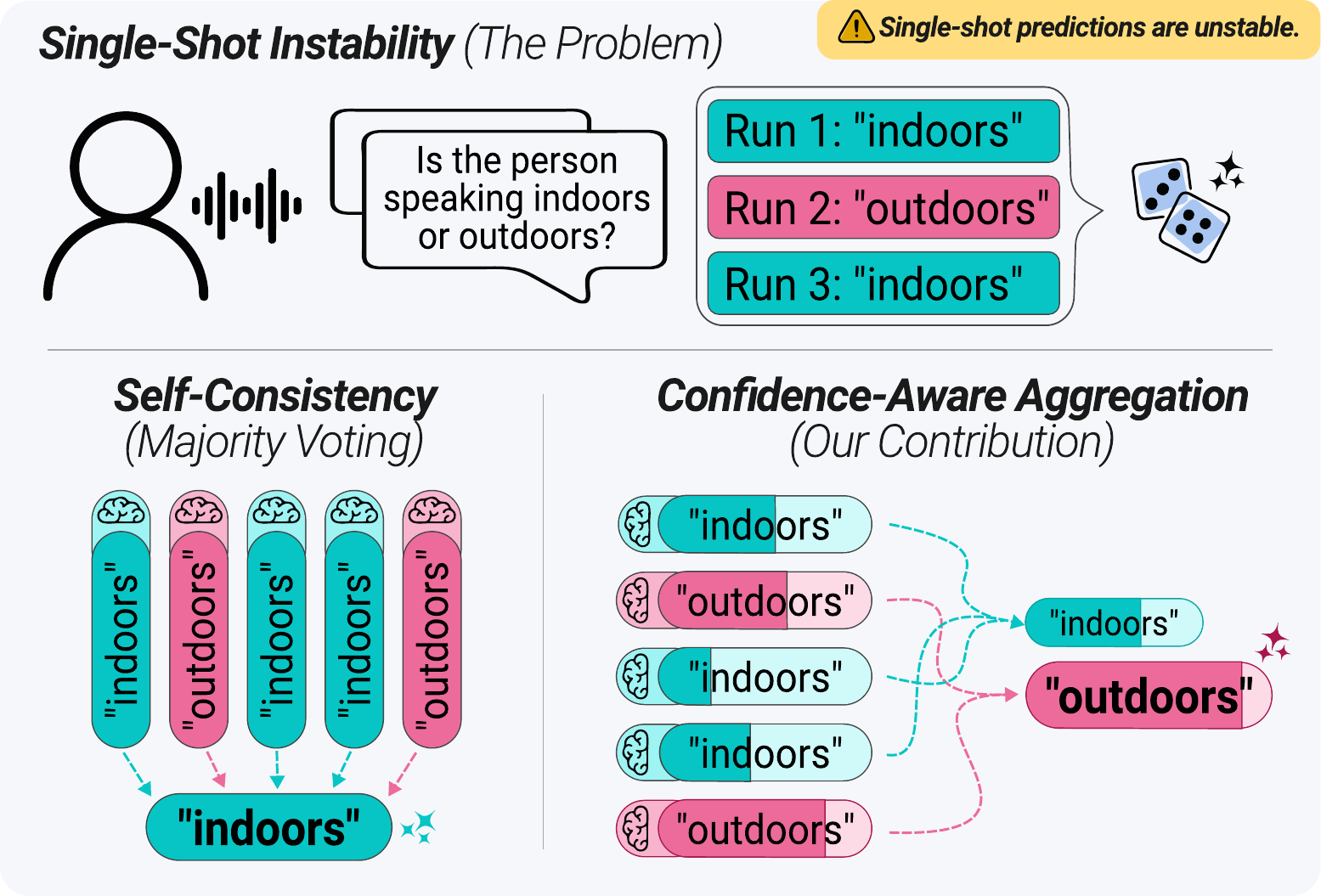}
  \caption{Variability in LALMs leads to inconsistent single-shot predictions. Self-consistency improves accuracy but assumes equal reliability. Leveraging confidence scores enables principled aggregation, turning variability into robustness, calibration, and improved reasoning.}
  \label{fig:teaser_intro}
\end{figure}

A key driver of this fragility is output variability. The same model can yield different answers to an identical audio–question pair under minimal decoding or prompting changes, reflecting both epistemic uncertainty from incomplete evidence and aleatoric uncertainty from intrinsically ambiguous signals (\cref{fig:teaser_intro}). Rather than treating variability as a liability, in this paper we ask: \textit{can LALMs exploit it to produce more accurate, robust, and well-calibrated answers?}

In Large Language Models (LLMs), variability has been successfully leveraged via self-consistency: multiple reasoning paths are sampled and aggregated to mitigate spurious steps and overconfident errors \cite{wang2023selfconsistency, shinn2023reflexion, zelikman2022star}. Initially introduced for arithmetic and commonsense reasoning \cite{wei2022chain, zelikman2022star2}, subsequent variants improve aggregation by accounting for heterogeneous reasoning trajectories without additional training or parameter updates \cite{huang2025efficient, chen2024universal, wang2025ranked, zhang2024earlystop}, showing that variability can be transformed into reliability. Recent work further highlights that not all reasoning paths are equally informative, motivating confidence-aware self-consistency, which leverages the model’s own self-certainty to score outputs, using token-level likelihoods as proxies for epistemic uncertainty \cite{xiong2024confidence, huang2024adaptive, kim2025calibrated}.

Transferring self-consistency from text to audio is non-trivial. Audio reasoning traces are conditioned on temporally structured, acoustically ambiguous signals with overlapping sources and weakly localized evidence, such that perceptual errors---for example, misattributing an onset or conflating overlapping sources---propagate into the reasoning trace itself instead of remaining confined to a linguistic inference step. This behavior challenges two implicit assumptions underlying text-based self-consistency: that sampled reasoning paths are approximately independent, and that token-level log-probabilities are calibrated proxies for correctness. Under these conditions, there is no guarantee that self-certainty signals remain meaningful or reliably reflect the correctness of a reasoning path.

Extending these ideas beyond text, inference-time self-consistency has begun to appear in multimodal reasoning, primarily for vision-language models\cite{faria2025selfgeosense, bansal2025honeybee, lin2025inference, nguyen2024selfensemble}. These approaches typically rely on surface-level majority voting or external reward-based re-ranking \cite{ko2025flexjudge, gunjal2024haldetect, lin2025hearing}. This pattern extends to audio reasoning, with Audio-CoT representing the first study applying  majority-vote aggregation to multiple sampled reasoning paths \cite{ma2025audio}. Whether a model's own confidence in each candidate can be exploited to distinguish among sampled reasoning paths and improve upon uniform voting remains an open question. This idea echoes a close analogue in speech processing, where second-pass rescoring and re-ranking of N-best ASR hypotheses using auxiliary language-model scores has been studied extensively, for example with semantic auxiliary tasks \cite{yang2021multi}. 

In this work, we investigate whether LALMs can improve Audio Question Answering (AQA) via inference-time aggregation strategies that jointly exploit output diversity and model self-certainty. Specifically, we score each sampled response with a log-probability-based confidence measure and compare five selection and aggregation rules spanning three conceptual regimes: extremal selection (Best-of-$N$), uniform consensus (Majority Vote), and confidence-weighted additive aggregation (Borda voting and Log-Sum-Exp voting). Our contributions are threefold: (1) the first systematic study of context-aware aggregation strategies for LALMs on multi-step reasoning AQA; (2) a log-probability-based scoring pipeline that consistently outperforms random sampling and vanilla majority voting; and (3) a comprehensive analysis of how each aggregation strategy scales with sampling budget, characterizing accuracy and calibration gains as the budget grows and how these gains depend on the level of agreement among sampled responses.

\section{Methodology}
\label{sec:methodology}

\subsection{Audio Question Answering as conditional generation}

Let $\mathcal{X}$ denote the space of audio inputs and $\mathcal{Q}$ the space of natural-language questions.  
An audio question answering (AQA) instance is a pair $(x,q) \in \mathcal{X} \times \mathcal{Q}$, with the goal of predicting a short answer $a \in \mathcal{Y}$, where $\mathcal{Y}$ is the set of valid answer strings.

A large audio language model (LALM) with fixed parameters $\theta$ defines a conditional distribution over token sequences:
\begin{equation}
p_\theta(y \mid x, q),
\end{equation}
where $y = (y_1, \dots, y_T)$ is generated autoregressively given $(x,q)$.  
Typically, the model output $y$ comprises a \emph{reasoning trace} followed by a \emph{final answer}.  
We denote the mapping from the full sequence to the normalized answer as
\begin{equation}
a = \mathrm{ans}(y) \in \mathcal{Y}.
\end{equation}

Because decoding is stochastic (e.g., temperature sampling), repeated inference on the same $(x,q)$ may yield different outputs.  
This motivates aggregation strategies that leverage output diversity to improve reliability and calibration.

\subsection{Answer sampling}

For each input $(x,q)$, we draw $N$ decoding samples:
\begin{equation}
y^{(i)} \sim p_\theta(\cdot \mid x, q), 
\qquad i = 1, \dots, N,
\end{equation}
and extract the corresponding answers:
$a^{(i)} = \mathrm{ans}(y^{(i)})$.

Let $\mathcal{A}_N = \{ a^{(1)}, \dots, a^{(N)} \}$ be the multiset of sampled answers.  
Increasing $N$ empirically improves coverage of plausible reasoning trajectories, though samples are not strictly independent.

\subsection{Self-certainty scoring}

To quantify model confidence, we compute the conditional log-likelihood of each generated sequence:
\begin{equation}
\log p_\theta(y^{(i)} \mid x, q)
=
\sum_{t=1}^{T_i} \log p_\theta(y^{(i)}_t \mid x, q, y^{(i)}_{<t}).
\label{eq:avg_logprob}
\end{equation}

In line with \cite{kang2025scalable}, we define a \emph{self-certainty score} $s^{(i)}$ for each sample.  
For generality, we formulate it over the full sequence (reasoning trace + answer):
\begin{equation}
s^{(i)} = \frac{1}{T_i} \log p_\theta(y^{(i)} \mid x, q),
\end{equation}
where $T_i = |y^{(i)}|$.  
Length normalization mitigates verbosity bias, ensuring that longer sequences do not receive higher scores artificially.

\subsection{Selection and aggregation strategies}
\label{sec:aggregation_strategies}

Let $\mathcal{A}$ denote the set of unique answers in $\mathcal{A}_N$, and define the index sets:
\begin{equation}
\mathcal{I}_a = \{ i \in \{1, \dots, N\} \mid a^{(i)} = a \}, \quad a \in \mathcal{A}.
\end{equation}

All aggregation rules are deterministic given the samples and self-certainty scores.  

\noindent\textbf{Best-of-$N$ (extremal selection).}  
Select the single most confident sample answer:
\begin{equation}
\hat{a}_{\mathrm{BoN}} = a^{(i^*)}, \quad
i^* = \arg\max_{i} s^{(i)}.
\end{equation}
This method prioritizes high-confidence outputs.

\noindent\textbf{Majority Vote (uniform consensus).}  
Select the answer appearing most frequently among all samples:
\begin{equation}
\hat{a}_{\mathrm{maj}} = \arg\max_{a \in \mathcal{A}} |\mathcal{I}_a|,
\end{equation}
where $|\mathcal{I}_a|$ counts how many times $a$ occurs. Here, each sample contributes equally, ignoring confidence. This captures consensus among reasoning trajectories.
\\\\
\noindent\textbf{Borda (rank-weighted aggregation).}  
Rank samples by decreasing self-certainty, $\mathrm{rank}(i) \in \{1, \dots, N\}$ (rank 1 = highest), and assign power-weighted scores:
\begin{equation}
w_i = (N - \mathrm{rank}(i) + 1)^\alpha, \quad \alpha > 0,
\end{equation}
and aggregate per answer:
\begin{equation}
V_{\mathrm{pBorda}}(a) = \sum_{i \in \mathcal{I}_a} w_i, \quad
\hat{a}_{\mathrm{pBorda}} = \arg\max_{a \in \mathcal{A}} V_{\mathrm{pBorda}}(a).
\end{equation}
This smoothly interpolates between uniform majority voting ($\alpha \to 0$) and extremal selection ($\alpha \to \infty$).
\\ \\
\noindent\textbf{Soft Log-Sum-Exp} (SoftLSE). Aggregate answers using a temperature-controlled softmax over self-certainty scores:
\begin{equation}
V_{\tau}(a) = \tau \log \sum_{i \in \mathcal{I}_a} \exp\left(\frac{s^{(i)}}{\tau}\right), 
\quad
\hat{a}_{\tau} = \arg\max_{a \in \mathcal{A}} V_{\tau}(a),
\end{equation}
Special cases include $\tau = 1$: standard Log-Sum-Exp (LSE) and the limit $\tau \to 0^+$, which recovers extremal selection.

\noindent\textbf{Summary}
For each input $(x,q)$, the inference procedure consists of three steps: (1) \textit{Sampling}: Generate $N$ sequences $y^{(i)}$, extract answers $a^{(i)}$;  
(2) \textit{scoring}: Compute self-certainty scores $s^{(i)}$ for each sample; and (3) \textit{Aggregation}: Combine $\{(a^{(i)}, s^{(i)})\}_{i=1}^N$ deterministically using strategies such as extremal selection, majority vote, or confidence-weighted aggregation to produce a final prediction $\hat{a}$.

\section{Experimental Setup}
This section presents our experimental protocol for evaluating confidence-aware selection and aggregation in multi-step reasoning Audio Question Answering (AQA). We focus solely on inference-time strategies, without any additional training or model adaptation.

\subsection{Dataset and models}
\textbf{MMAR dataset}. MMAR \cite{ma2025mmar, ma2026interspeech2026audioreasoning} is a dataset for audio reasoning comprising 1,000 audio–question–answer triplets from real-world videos. Each item is a multiple-choice question with fixed options. The benchmark covers seven content categories, including general sound, music, speech, and mixtures thereof, and follows a four-level taxonomy (Signal, Perception, Semantic, and Cultural).

\noindent\textbf{Models.} We evaluate four audio-language foundation models: Qwen3-Omni-Thinking and Qwen3-Omni-Instruct \cite{xu2025qwen3omni} (30B total / 3B active parameters), Qwen2-Audio \cite{Qwen2-Audio} ($8.4$B parameters) and Audio Flamingo 3 \cite{ghosh2025audio} (built on a $7$B backbone).

\begin{figure*}[t]
  \centering
  \includegraphics[width=\linewidth]{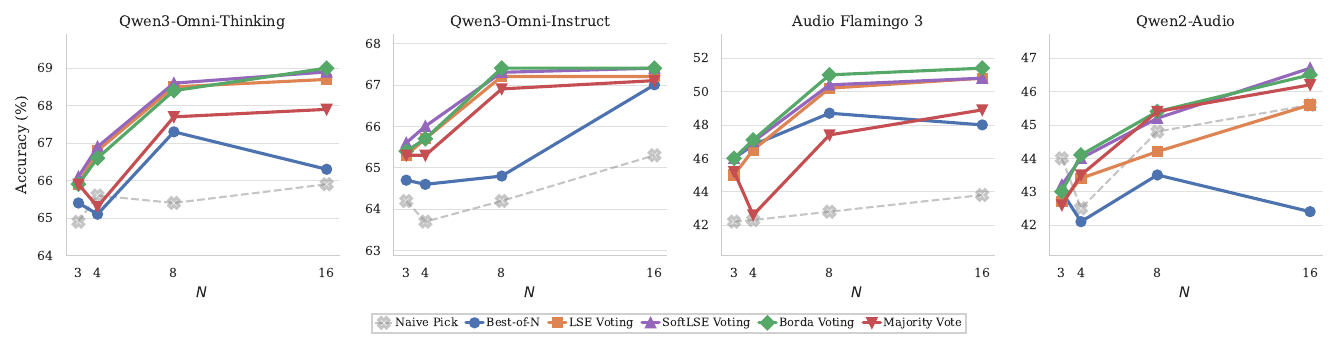}
  \caption{Accuracy (\%) vs.\ sampling budget $N$ for four large audio-language models. Confidence-aware methods consistently outperform the Naive Pick baseline, and achieve typically higher accuracy than Majority Vote with the accuracy gap widening as $N$ increases.}
  \label{fig:per_model_acc_vs_N}
\end{figure*}

\subsection{Sampling protocol}
For each audio-question pair, we generate $N \in \{3,4,8,16\}$ candidate responses using a fixed set of $N$ decoding temperatures, sampled uniformly from $[0.1, 1]$ once per inference run and reused across all audio-question pairs. These responses form the self-consistency candidate pool. We adopt this setup because it is designed to surface the broader epistemic diversity that confidence-aware aggregation is meant to exploit, whereas fixed-temperature sampling constrains the token distribution and would likely understate the benefit of self-certainty weighting. To isolate aggregation-driven improvements from inherent sampling variance, we establish a \textit{Naive Pick} baseline, defined as a uniform random selection $a^{(i)}$ from the candidate pool ($i \sim \mathcal{U}(1, N)$). This distinguishes the model's average stochastic performance from the benefits of structured consolidation of evidence.

\subsection{Response parsing and filtering}
\label{sec:response_parsing_filtering}
We first obtain raw generations from all models. For Qwen2-Audio and Audio Flamingo 3, we apply a deterministic LLaMA 3.1 8B-Instruct parser to extract answer spans from outputs that do not strictly follow the required format. The parser operates only on the raw text, has no access to the audio input or self-certainty scores, and does not generate or revise answer content \cite{yue2024mmmu}. We then filter, for all models, the extracted responses by validating them against the fixed set of choices for each item: after normalization, a response is retained only if it is non-empty and matches exactly one choice; unmatched or empty responses are discarded. Aggregation (\cref{sec:aggregation_strategies}) is performed exclusively over these validated answers.


\subsection{Self-certainty scoring and aggregation}
Following \cite{kang2025scalable}, we define self-certainty as the model’s conditional likelihood of its generated output given the joint audio–text input, as expressed in \cref{eq:avg_logprob}. This formulation provides a principled measure of the model’s own confidence in its predictions. In our setup, we compute self-certainty scores over the concatenation of both the parsed reasoning trace and the final answer produced by the model. This ensures that the confidence estimate reflects both the correctness of the answer and the internal reasoning process.
We organize the five inference-time selection and aggregation methods introduced in \cref{sec:aggregation_strategies} into three conceptual geometries: extremal (Self-Certainty), uniform consensus (Majority Vote), and confidence-weighted additive (SoftLSE and Borda). For SoftLSE and Borda, the temperature and scaling parameters $\alpha$ and $\tau$ are selected through grid search over $[0,4]$ with increments of $0.25$.

\subsection{Evaluation metrics}
We evaluate accuracy on the MMAR benchmark and report three complementary metrics to characterize model behavior beyond raw accuracy. \textbf{Answer rate.} We report the fraction of valid, parseable responses before and after response filtering (\cref{sec:response_parsing_filtering}), to quantify how filtering affects the usable sample pool. \textbf{Calibration (ECE).} We compute Expected Calibration Error~\cite{guo2017calibration} (ECE, 10 bins) over the valid (post-filtered) response pool of size $N$. ECE requires a single confidence value $c(\hat{a})$ per aggregated answer $\hat{a}$, which we define per method as follows: Best-of-$N$ uses the exponentiated score of the selected sample, $c(\hat{a}_{\text{BoN}}) = \exp(s^{(i^*)})$; Majority Vote uses the winning vote fraction, $c(\hat{a}_{\text{maj}}) = |I_{\hat{a}}| / N$; Borda uses the winning answer's normalized weight, $c(\hat{a}_{\text{pBorda}}) = V_{\text{pBorda}}(\hat{a}) / \sum_{a \in \mathcal{A}} V_{\text{pBorda}}(a)$; and LSE/SoftLSE use a softmax over $V_\tau(a)$ across candidate answers,
\begin{equation}
P(a) = \frac{\exp(V_\tau(a))}{\sum_{a' \in \mathcal{A}} \exp(V_\tau(a'))}, \qquad
c(\hat{a}_\tau) = \max_{a \in \mathcal{A}} P(a).
\label{eq:ece_conf}
\end{equation}

\noindent\textbf{Accuracy by agreement level.} To assess the impact of sample variability, we group instances by agreement among the $N=16$ sampled responses (the highest sampling budget) into three levels: high consensus ($\mathrm{Agr} \geq 0.75$), medium agreement ($0.50 \leq \mathrm{Agr} < 0.75$), and high disagreement ($\mathrm{Agr} < 0.50$)~\cite{dietterich2000ensemble}, where $\mathrm{Agr} = \max_{a \in \mathcal{A}} |I_a|/N$ is the fraction of sampled responses agreeing with the modal answer. For each level, we report absolute accuracy gains over Majority Vote in percentage points (pp).

\section{Results and discussion}
We report the outcomes of our experiments, shedding light on the impact of confidence-aware selection and aggregation strategies and their role in model robustness and calibration.

\subsection{Effect of response filtering}
\Cref{tab:answer_rate_accuracy_effect} quantifies the impact of validity filtering on answer rate and downstream accuracy, averaged across $N \in \{3, 4, 8, 16\}$ and all selection and aggregation methods (excluding Naive Pick). Here, \textit{Unfiltered} includes all parsed answer spans, whereas \textit{Valid} retains only responses that are non-empty and exactly match one of the provided choices, as defined in \cref{sec:response_parsing_filtering}. Validity filtering consistently increases the answer rate across the four evaluated models, with the largest gains observed for Audio Flamingo 3 (+5.26 pp) and Qwen3-Omni-Instruct (+4.60 pp). By contrast, Qwen3-Omni-Thinking is already near ceiling (98.79\%) and shows only marginal improvement (+0.84 pp). Accuracy follows a similar pattern, with larger gains for models with lower initial answer rates. Filtering effectively removes responses that cannot be associated with a valid choice and ensures that aggregation operates exclusively on valid answers. All subsequent analyses consider valid responses only.

\begin{table}[htbp]
\centering
\footnotesize
\caption{Effect of validity filtering on answer rate and accuracy for the four evaluated model variants, averaged over $N \in \{3,4,8,16\}$ and five selection methods (Naive Pick excluded). $\Delta$ AR and $\Delta$ Acc.~denote the changes in answer rate and accuracy, respectively, in percentage points.}
\label{tab:answer_rate_accuracy_effect}
\begin{tabular}{l rrr r}
\toprule
& \multicolumn{3}{c}{\textbf{Answer Rate (\%)}} & \textbf{Accuracy} \\
\cmidrule(lr){2-4} \cmidrule(lr){5-5}
\textbf{Model} & Unfiltered & Valid & $\Delta$ AR & $\Delta$ Acc \\
\midrule
  Qwen3-Omni-Thinking & 98.79 & 99.62 & +0.84 & +0.38 \\
  Qwen3-Omni-Instruct & 94.00 & 98.60 & +4.60 & +1.90 \\
  Audio Flamingo 3 & 94.21 & 99.48 & +5.26 & +1.74 \\
  Qwen2-Audio & 94.42 & 97.22 & +2.80 & +1.20 \\

\bottomrule
\end{tabular}
\end{table}

\subsection{Overall performance}

\Cref{fig:per_model_acc_vs_N} reports accuracy as a function of sampling budget $N$ for each model. For both Qwen3-Omni variants and Audio Flamingo 3, all aggregation strategies  (Borda, LSE, SoftLSE) outperform Naive Pick, and further improve over Majority Vote. This shows that self-certainty provides useful signal about answer correctness, but is most effective when aggregated across the sampled pool: Best-of-$N$ consistently underperforms Borda, LSE, and SoftLSE, indicating that confidence is more useful as a weighting signal than for selecting the most confident sample. For Qwen2-Audio, aggregation yields only marginal gains over Majority Vote, likely reflecting limitations in its architecture. We nevertheless retain it for subsequent analyses. Unless noted otherwise, results are aggregated across models.

\subsection{Scaling behavior with sampling budget}
\label{sec:scaling}

\begin{figure}[t]
  \centering
  \includegraphics[width=\linewidth]{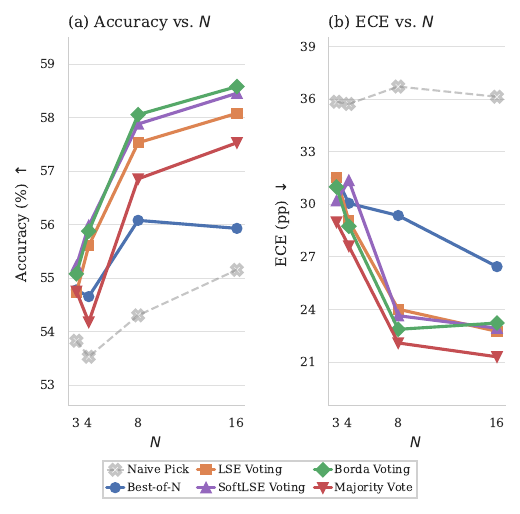}
  \caption{Accuracy (a) and ECE (b) versus sampling budget $N$, averaged across audio-language models. Additive aggregation scales best, improving accuracy and calibration, while Self-Certainty and Naive Pick show limited gains; improvements saturate beyond $N{=}8$.}
  \label{fig:accuracy_vs_sampling_budget}
\end{figure}

\Cref{fig:accuracy_vs_sampling_budget} shows accuracy and calibration (ECE) as a function of sampling budget $N$, averaged across all evaluated models, with \cref{tab:scaling_delta} summarizing the total change from $N{=}3$ to $N{=}16$.

All methods improve with increasing $N$, but magnitudes differ. Additive aggregation methods (Borda, LSE, SoftLSE) show the largest gains, ranging from 3.2 to 3.5 points, followed by Majority Vote (+2.8 points), while extremal selection (Best-of-$N$) and the Naive Pick baseline improve by only 1 point.

Calibration trends mirror accuracy. Naive Pick remains miscalibrated, whereas all aggregation methods reduce ECE by 7.26 to 8.77 points. Consolidating evidence improves reliability, in line with recent calibrated audio reasoning works \cite{gibier2025audio}. Additive methods and majority voting achieve the largest gains; extremal selection shows limited improvement, an observation consistent with \cite{kang2025scalable}, which found that response quality initially rises as confidence decreases but later deteriorates sharply.

Overall, increasing the sampling budget $N$ mainly benefits methods that consolidate evidence. Gains saturate beyond $N=8$, and the consistent scaling across architectures suggests a general property of stochastic decoding rather than model-specific effects.


\begin{table}[t]
\centering
\caption{Inference-time scaling gains ($N=3 \to 16$). $\Delta$Acc (pp) and $\Delta$ECE (pp) averaged across models.}
\label{tab:scaling_delta}
\begin{tabular}{lcc}
\toprule
Aggregation Method & $\Delta$ Accuracy (pp) & $\Delta$ ECE (pp) \\
\midrule
Naive Pick      & +1.32 & +0.03 \\
Best-of-$N$     & +1.11 & -5.06 \\
Majority Vote   & +2.77 & -7.69 \\
LSE Voting      & +3.35 & -8.77 \\
SoftLSE Voting  & +3.23 & -7.26 \\
Borda Voting    & +3.50 & -7.76 \\
\bottomrule
\end{tabular}
\end{table}

\subsection{Analyzing accuracy by agreement level}

\begin{figure}[t]
  \centering
  \includegraphics[width=\linewidth]{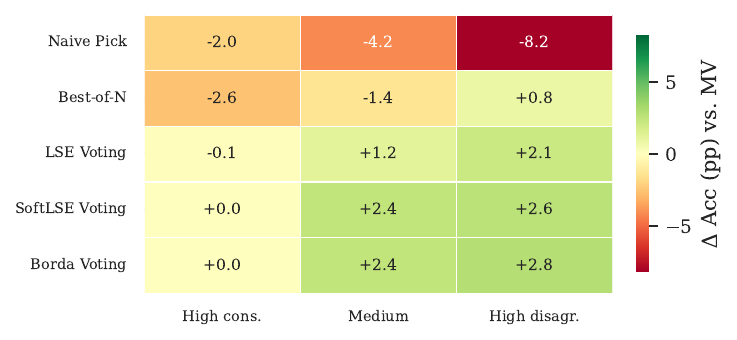}
\caption{Accuracy gain (pp) over Majority Vote for $N=16$, averaged across models. Confidence-aware aggregation yields largest improvements under medium and high disagreement, with negligible gains in high-consensus cases.}
  \label{fig:cross_model_variability}
\end{figure}


\Cref{fig:cross_model_variability} reports accuracy by agreement level for each selection and aggregation strategy relative to Majority Vote. As a reference, Majority Vote achieves approximately 67\% accuracy in the high-consensus regime, decreasing to approximately 43\% in the medium-agreement regime and 32\% in the high-disagreement regime. Two patterns emerge. Firstly, the benefit of confidence-aware aggregation grows as agreement decreases. In the high-consensus regime, methods largely converge because the sampled responses already provide a strong consensus, leaving limited scope for reweighting to improve the decision.  In contrast, under medium and high disagreement, self-certainty-based aggregation methods (LSE, SoftLSE, and Borda) outperform Majority Vote by 1.2--2.4~pp and 2.1--2.8~pp, respectively, suggesting that self-certainty helps identify more reliable responses when predictions conflict. Secondly, these gains emerge precisely where Majority Vote becomes less reliable, indicating that aggregating evidence from self-certainty signals can better exploit informative responses than treating every response as equally trustworthy. Overall, confidence-aware aggregation strategies appear particularly valuable for ambiguous or conflicting audio-question instances, where vanilla response agreement alone provides a weaker basis for aggregation.

\section{Conclusion}
This study addressed whether the output variability that undermines single-shot LALM predictions could instead be exploited to improve accuracy and calibration. We proposed an inference-time framework that samples multiple candidate responses and aggregates them using self-certainty, comparing Best-of-$N$, Majority Vote and confidence-weighted aggregation (Borda, LSE, SoftLSE) across four LALMs on the MMAR dataset. We showed that aggregation strategies outperformed Majority Vote, with the gains being most pronounced where consensus is least reliable. Notably, Best-of-$N$ underperformed aggregation strategies, indicating that confidence is more valuable as a weighting signal than as a selection criterion. Overall, our results show that audio reasoning robustness can be improved at test time by leveraging the model's own confidence without additional training. Validation on further benchmarks and exploration of audio-specific uncertainty signals remain open directions.

\clearpage
\section{Acknowledgment}
This work was supported by Hi!Paris, State funding managed by the French National Research Agency (ANR) under the France 2030 program, reference ANR-23-IACL-0005, and by the European Union (ERC, HI-Audio, 101052978). Views and opinions expressed are however those of the author(s) only and do not necessarily reflect those of the European Union or the European Research Council. Neither the European Union nor the granting authority can be held responsible for them.

\bibliographystyle{IEEEtran}
\bibliography{refs}

\end{document}